\documentclass[twocolumn,aps,prc,showpacs,superscriptaddress,preprintnumbers,floatfix,nofootinbib]{revtex4}
\usepackage{epsfig,graphics}
\usepackage{graphicx}
\usepackage{dcolumn}
\usepackage{bm}
\usepackage{amsmath}
\usepackage[usenames]{color}
\usepackage{ulem} 

\newcommand{ \srt  }{$\sqrt{s_{_{\rm NN}}}$}

\newcommand{ \pt } {${p_{T}}$}

\usepackage{CJK}
\usepackage{epstopdf}
\hfuzz=\maxdimen
\begin{document}
\begin{CJK*}{GBK}{song}

\title{$\Omega$ and $\phi$ in Au + Au ~collisions at \srt~= 200 and 11.5 GeV from a multiphase transport model}
\author{Y. J. Ye}
\affiliation{Shanghai Institute of Applied Physics, Chinese Academy of Sciences, Shanghai 201800, China}
\affiliation{University of Chinese Academy of Sciences, Beijing 100049, China}
\author{J. H. Chen}
\email{chenjinhui@sinap.ac.cn}
\affiliation{Shanghai Institute of Applied Physics, Chinese Academy of Sciences, Shanghai 201800, China}
\author{Y. G. Ma}
\email{ygma@sinap.ac.cn}
\affiliation{Shanghai Institute of Applied Physics, Chinese Academy of Sciences, Shanghai 201800, China}
\affiliation{ShanghaiTech University, Shanghai 200031, China}
\author{S. Zhang}
\affiliation{Shanghai Institute of Applied Physics, Chinese Academy of Sciences, Shanghai 201800, China}
\author{C. Zhong}
\affiliation{Shanghai Institute of Applied Physics, Chinese Academy of Sciences, Shanghai 201800, China}

\date{\today}

\begin{abstract}

Within the framework of a multiphase transport model, we study the production and properties of $\Omega$ and $\phi$ in Au + Au collisions with a new set of parameters for $\sqrt{s_{NN}}$ = 200 GeV and with the original set of parameters for $\sqrt{s_{NN}}$ = 11.5 GeV, respectively. The AMPT model with the string melting version provides a reasonable description at $\sqrt{s_{NN}}$ = 200 GeV and while the AMPT model with default version describes the data well at $\sqrt{s_{NN}}$ = 11.5 GeV. It indicates that the system created at top RHIC energy is dominated by partonic interaction and while the hadronic interaction becomes important at lower beam energy, such as $\sqrt{s_{NN}}$ = 11.5 GeV. The comparison of $N(\Omega ^++\Omega^-)/[2N(\phi)]$ ratio between data and calculations further supports the argument. Our calculations can generally describe the data of nuclear modification factor as well as elliptic flow.
\end{abstract}


\maketitle

\section{Introduction}

Lattice QCD calculations predict a critical point which separates a first-order phase transition and a smooth crossover in the QCD phase diagram~\cite{phase1,phase2}. Experimental efforts from the Relativistic Heavy Ion Collider (RHIC) at BNL have implemented to dedicate  the Beam Energy Scan  program. Exciting results of QCD matter including antimatter nuclei detection and its interaction measurement \cite{Science,Nature,Nature2,Ma_review,Zhang_NST},
non-monotonic behaviour of conserved quantity fluctuation~\cite{STARnetp,STARnetq,STARpidC,Ko2016,Luo} and strangeness enhancement~\cite{NA491,NA492} etc have displayed many interesting features. While,  the chiral electro-magnetic effects demonstrated  rich phenomena for QCD matter in relativistic heavy-ion collisions  \cite{Dima,HuangXG,MaGL}.
Also,  the increasing difference between the elliptic flows of particles and antiparticles, thus a breaking of the
number of constituent quark scaling of elliptic flow has been observed as the collision energy decreases \cite{STAR_v2} and might be related to phase transition \cite{Tian,Xu_PRL,Ko2013}.
Among these measurements, one of the new ideas is to study the $\Omega$/$\phi$ ratio in a broad energy range to search for phase transition signal from partonic degree of freedom (d.o.f) dominated system to hadronic interaction dominated system~\cite{Omegaphical}.  Transverse momentum ($p_T$) dependent  $\Omega$/$\phi$ ratio was observed to  fall on a consistent trend at high collision energies, but starts to deviate in peripheral collisions at \srt ~= 19.6, 27, and 39 GeV, and in central collisions at 11.5 GeV in the intermediate $p_T$ region of $2.4-3.6$ GeV/$c$ \cite{OmegaphiSTAR}  by the RHIC-STAR collaboration\cite{STAR_detector,XuYF,MaLong}. Also, the number of constituent quark scaled $\Omega$/$\phi$ ratios show a suppression of strange quark production in central collision at \srt ~= 11.5 GeV in comparison with those at \srt ~$\geq$ 19.6 GeV~\cite{OmegaphiSTAR}. The data indicates that there is likely a change of the underlying strange quark dynamics in the transition from quark matter to hadronic matter at collision energies below 19.6 GeV, however, the understanding on detailed transport dynamics is still missing. In this paper, we study the  production of $\Omega$ and $\phi$ at $\sqrt{s_{NN}}$ = 200 and 11.5 GeV by a multiphase transport (AMPT) model with new tuned input parameters.

The article is organized as follows. In Sec. II, we introduce the AMPT model in brief and discuss the parameters in the model which are relevant to the $p_T$ spectra shape of identified particle. We present the updated calculation on $\Omega$ and $\phi$ spectra, the particle ratio, the nuclear modification factor as well as the elliptic flow in Sec. III. Finally, a summary is given in Sec. IV.

\section{the ampt model}

The AMPT model has been used extensively to describe the dynamics of system evolution created in high energy heavy-ion collisions~\cite{AMPT-model}. The default version of AMPT model contained four parts: the fluctuating initial conditions from the HIJING model~\cite{HIJING}, the elastic parton cascade from ZPC~\cite{ZPC} for minijet partons, the Lund string model~\cite{Lund} for hadronization, and the ART~\cite{ART} for hadron transport. It describes the rapidity distributions and transverse momentum spectra in heavy-ion collisions from SPS to RHIC energies reasonably well but underestimates the elliptic flow observed at RHIC~\cite{AMPT-model}. It was pointed out by authors of the model that most of the energy produced in the overlapping volume of heavy-ion collisions are in hadronic strings and thus not included in the parton cascade in the default AMPT model~\cite{Lin2002}. And the string melting version of the AMPT was developed, where all excited hadronic strings in the overlap volume are converted into partons~\cite{Lin2002}. The AMPT model with the string melting scenario consists of fluctuating initial conditions from the HIJING model~\cite{HIJING}, the elastic parton cascade ZPC~\cite{ZPC} for all partons from the melting of hadronic strings, a quark coalescence model based on the quark spatial information for hadronization, and the ART hadronic cascade~\cite{ART}. Since all hadronic strings in HIJING are converted to partons in the melting version, the parton density in the ZPC is quite dense and it can reasonably fit the elliptic flow in heavy-ion collisions at RHIC energy~\cite{AMPT-model}. However, it fails to reproduce well the rapidity distributions and the $p_T$ spectra when  the same Lund $a$ and $b$ parameters are used as in the default version.

In previous studies, it was found that the multiplicity of charged particles measured in heavy ion collision at RHIC energy and at LHC energy could be well described by the melting AMPT model with modified parameters of $\it {a}$ and $\it b$ in the Lund string fragmentation function ~\cite{ab,ablin1}

    \begin{equation}
        f(z)\propto z^{-1}(1-z)^{a}exp(-bm_{\perp }^{2}/z),
        \label{dn_Lund}
    \end{equation}
where $\it z$ is the light-cone momentum fraction of the produced hadron with respect to that of the fragmenting string. The average square transverse momentum is then given by~\cite{AMPT-model}
     \begin{equation}
        <p_{\perp }^{2}>=\frac{\int p_{\perp }^{2}f(z)d^{2}p_{\perp }dz}{\int f(z)d^{2}p_{\perp }dz}.
        \label{dn_Lund1}
    \end{equation}
For massless particles, it reduces to~\cite{AMPT-model}
     \begin{equation}
        <p_{\perp }^{2}>=\frac{1}{b(2+a)}.
        \label{dn_Lund2}
    \end{equation}
Recently, a systematic study of predictions for \srt~=5.02 TeV Pb+Pb collisions is updated by the same authors with new tuned parameters~\cite{ablin2}. Here we tune the parameters in the model to study the dynamics of $\Omega$ and $\phi$ production in Au+Au collisions at \srt = 200 and 11.5 GeV.

    \begin{table}[htbp]
       \caption{Values of $\bm a$ and $\bm b$ in the Lund string fragmentation function and $\alpha _{s}$, $\mu$ relevant to the parton scattering cross section via $\sigma\approx 9\pi\alpha _{s}^{2}/\left ( 2\mu^{2} \right )$. Set A \cite{AMPT-model} and Set B \cite{ab} are from previous studies, and Set C is the present work.}
       \label{para}
       \centering
       \begin{tabular}{p{45pt} p{45pt} p{45pt} p{45pt} p{45pt}}
          \hline
          \hline
             & $\bm a$ & $\bm b(GeV^{-2})$& $\alpha _{s}$ & $\mu(fm^{-1})$  \\
          \hline
          A  & 2.2 & 0.5 & 0.47 &   1.8   \\
          B & 0.5 & 0.9 & 0.33 & 3.2 \\
          C  & 0.55 & 0.15  & 0.33 & 3.2    \\
          \hline
          \hline
       \end{tabular}
       \end{table}

Table \ref{para} lists the parameters used in previous studies and the present work. The parameter set B provides better description of transverse momentum spectra of charged particle at RHIC energy than the set A parameter does   for the data of p+p and d + Au \cite{dAu}, though both give a softer spectrum in comparison with experimental data. We here tune the parameter again to try to match the data. In particular, we apply a small parton cascade cross section of 1.5 mb in comparison with the parameter sets of 3 mb as discussed in Ref.~\cite{ablin2}. According to the Eq.(3), we know that the parameters of $\it a$ and $\it b$ determine the \pt ~distribution of particle production. A large $\it a$ and $\it b$ will provide a small average square transverse momentum distribution then it shows us a sharp \pt ~spectrum and a small one will give a flat distribution. In the study of Lund string fragmentation model in the AMPT~\cite{AMPT-model-a}, we also know that quark-antiquark pair production probability is proportional to $exp(-\pi m_{\perp }^{2}/\kappa )$, where $\kappa$ is the string tension and $\kappa\propto \frac{1}{b(a+2)}$. Due to its large mass, strange quark production is suppressed by the factor $exp(-\pi (m_{s}^{2}-m_{u}^{2})/\kappa )$, compared to that of light quarks. So, larger values of $\it a$ and $\it b$ lead to a suppression of strangeness.

    \begin{figure}[htbp]
      \includegraphics[width=0.52\textwidth]{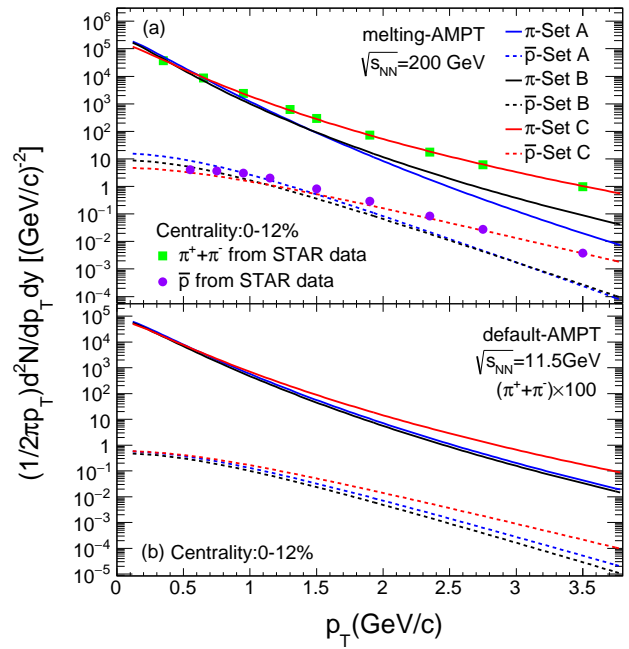}\\
      \caption{The $\pi$ and $\bar{p}$ invariant yields as a function of \pt~in Au+Au collision. (a)  the calculation of AMPT model with the string melting scenario for  $\sqrt{s_{NN}}$ = 200 GeV with different parameters of set A, set B and set C. The experimental data~\cite{star_compare} is also plotted for comparison; (b) the calculation from the default AMPT model at $\sqrt{s_{NN}}$ = 11.5 GeV.}
      \label{compare}
    \end{figure}

Figure \ref{compare} shows the transverse momentum  spectra of charged $\pi$ and $\bar{p}$ in Au+Au collisions at $\sqrt{s_{NN}}$ = 200 GeV from the string melting AMPT model (version 2.26t5 for this study, available online \cite{AMPT-Web}). The result at $\sqrt{s_{NN}}$ = 11.5 GeV from the default AMPT model is presented in panel (b). In panel (a), it is seen that all parameter sets reasonably describe the experimental data~\cite{star_compare} at low \pt. At high \pt, the parameter set C gives larger yields as a result of smaller energy loss of high \pt ~particles when the parton scattering cross section is smaller for parameter set C (1.5 mb) than parameter set A (10 mb). Parameter set C gives more flat \pt ~spectra in comparison with the parameter set A and B because of the smaller value of $\it a$ and $\it b$, which dominates charged particle \pt ~spectra as mentioned before. Parameter set C describes the experimental data well, and a smaller $\it b$ will reduce the suppression of strangeness, which relates to our study in this paper. So we choose the parameter set C at energy $\sqrt{s_{NN}}$ = 200 GeV, while keep the parameter set A at energy $\sqrt{s_{NN}}$ = 11.5 GeV in present work due to the strong suppression of strange quark production at lower energy, which will be illustrated by the strange matter \pt~spectra in the following discussion. Because the experimental data $\sqrt{s_{NN}}$ = 11.5 GeV is not published yet, panel (b) only shows our calculation, and the comparison with experimental data is not involved. Similar feature on transverse momentum spectrum for charge $\pi$ and $\bar{p}$ is observed in the AMPT model from three different sets of parameter as seen in the 200 GeV's (panel a).

\section{results}

\subsection{Production of $\Omega$ and $\phi$}

    \begin{figure}[htbp]
      \includegraphics[width=0.52\textwidth]{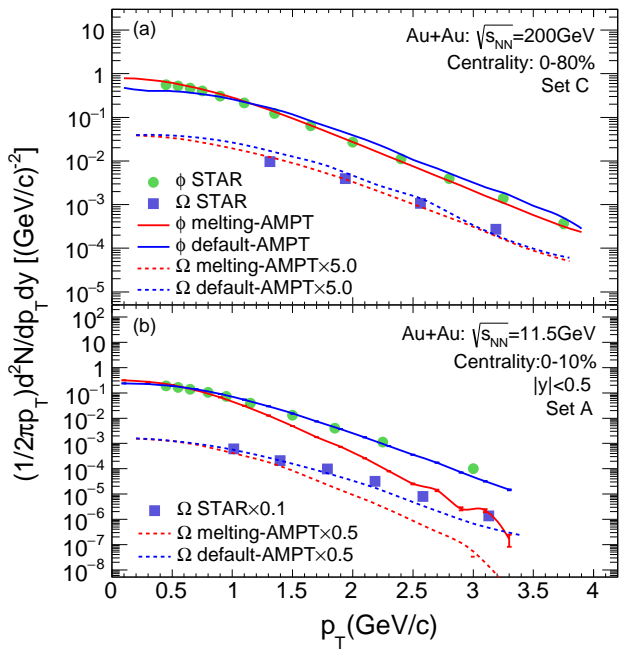}\\
      \caption{The invariant yields of $\Omega$ baryon and $\phi$ meson as a function of \pt~in Au+Au collisions. (a) presents the result at $\sqrt{s_{NN}}$ = 200 GeV and (b) is the result at  $\sqrt{s_{NN}}$ = 11.5 GeV. The calculations are compared with experimental data and different set of parameters are adopted.}
      \label{dndydpt200}
    \end{figure}
    
The $\Omega$ and $\phi$ invariant yields as a function of \pt~in Au+Au collisions from the AMTP model are shown in Figure \ref{dndydpt200}. The upper panel shows the yields of $\Omega$ and $\phi$ at $\sqrt{s_{NN}}$=200 GeV. For $\phi$ meson, the rapidity range is $|y|<1.0$, while for the $\Omega$, the rapidity range is $|y|<0.75$ as the same window in experimental analysis. The AMPT model with the string melting scenario describes the production of $\phi$ meson well at energy $\sqrt{s_{NN}}$ = 200 GeV\cite{star_phi}, while the default version gives a small deviation on the slope of the $\phi$ meson $p_T$ spectrum in comparison with data. Both AMPT with the string melting and the default version describes the slope of the $\Omega$ spectra and underpredicts the rate by a factor of 5.0~\cite{star_Omega}, which may be due to the current coalescence algorithm in the AMPT model, which leaves a space to further improve the algorithm or model in future~\cite{AMPT-model}.

Moving to the lower energy, as shown in the bottom panel of Fig. \ref{dndydpt200}, the default AMPT model describes the $\phi$ meson  data well~\cite{OmegaphiSTAR}, while the melting AMPT model describes the data at low \pt~but under predicts the production rate significantly at high \pt. This is probably due to the fact that the AMPT model with string melting scenario includes significant parton interaction which leads to a large energy loss when the partons pass through the medium before hadronization than the default AMPT. For the $\Omega$, the AMPT model with the default scenario describes the experimental data well after the yield of $\Omega$ is scaled by a factor 5.0. The production rate is strongly suppressed at high \pt~from the AMPT with the string melting scenario and deviated from the experimental data. The comparison between panel (a) and panel (b) indicates that parton interactions are important in Au + Au collisions at $\sqrt{s_{NN}}$ = 200 GeV while hadronic interaction becomes important at $\sqrt{s_{NN}}$ = 11.5 GeV. This is consistent with the findings on possible partonic dominated d.o.f to hadronic dominated d.o.f  from the RHIC BES  analysis for $\Omega$ and $\phi$ production~\cite{OmegaphiSTAR}.

\subsection{Strangeness dynamics in the system evolution}

A quark recombination model calculation by Hwa and Yang is established to describe the physics of the ratio of $\Omega$/$\phi$ invariant yields in Au + Au collisions at $\sqrt{s_{NN}}$=200 GeV~\cite{ratio}. In their model, if we only concern about the thermal contribution, the invariant yield distribution of $\Omega$ and $\phi$ is~\cite{ratio}:

    \begin{equation}
        \frac{dN_{\Omega}}{pdp}=g_{\Omega}C_{s}^{3}\frac{p^{2}}{27p_{0}}e^{-\frac{p}{T_s}},
        \label{dn_Omega}
    \end{equation}

    \begin{equation}
        \frac{dN_{\phi}}{pdp}=g_{\phi}C_s^{2}\frac{p}{4p_{0}}e^{-\frac{p}{T_s}},
        \label{dn_phi}
    \end{equation}
where $g_{\Omega}$ and $g_{\phi}$ are statistical factor, and $T_s$ is the inverse slope. $C_s$ is the normalization factor which is adjusted to fit the experimental data at low $p$, and has the dimension [momentum]$^{-1}$.
The $\Omega$/$\phi$ ratio $R_{\Omega/\phi}$ is then derived:
     \begin{equation}
        R_{\Omega/\phi}^{th}(p)=\frac{4g_{\Omega}C_{s}}{27g_{\phi}}p
        \label{ratio_op}
    \end{equation}
from equation~\ref{ratio_op}, one can find that the ratio raising linearly as momentum increases if one only considers the thermal contribution.

    \begin{figure}[htbp]
      \includegraphics[width=0.52\textwidth]{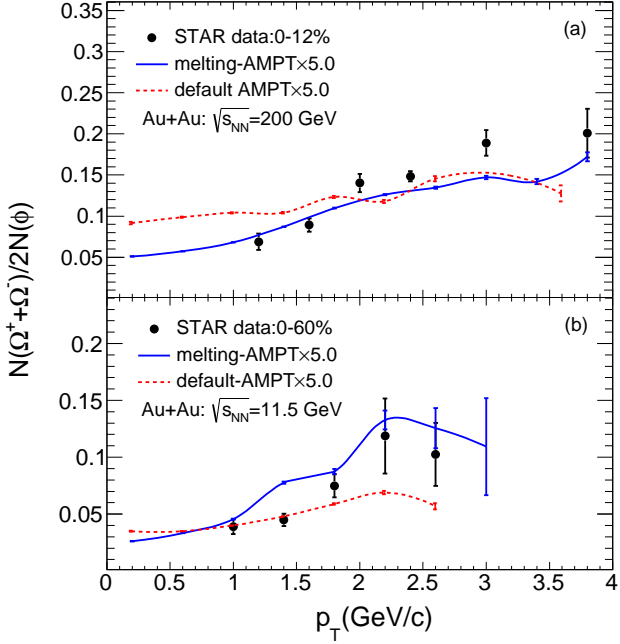}\\
      \caption{The ratio of $N(\Omega ^++\Omega^-)/[2N(\phi)]$ as a function of \pt ~in Au + Au collisions. (a) the AMPT  model calculation with the string melting version at $\sqrt{s_{NN}} =  200$ GeV (Parameter Set C) and (b)  the default AMPT model calculation at $\sqrt{s_{NN}} = 11.5$ GeV (Parameter Set A). }
      \label{ratio200}
    \end{figure}

Figure \ref{ratio200} shows the results from the AMPT model on the ratio of $N(\Omega ^++\Omega^-)/[2N(\phi)]$ in Au + Au collisions at mid-rapidity at $\sqrt{s_{NN}} = 200$ GeV and 11.5 GeV. In the upper panel, one can see that the ratio after the $\Omega$ yields scaled by a factor of 5.0 from the melting AMPT describes the experimental data~\cite{OmegaphiSTAR} reasonably well and gives us an approximately straight line, but the result from the default AMPT model doesn't describe the data. At $\sqrt{s_{NN}} = 11.5$ GeV, as shown in the bottom panel, both the default AMPT and the melting AMPT seem to describe the shape of the data with large statistical error. But as we discussed in previous session, the AMPT model with the string melting scenario under predicts the $\Omega$-baryon and $\phi$-meson spectra at high $p_T$ even though the ratios seem along the right way [c.f. Fig.~\ref{dndydpt200}]. We learnt that the ratio at $p_T$ = 0.0 from our calculation is higher than 0, contrary to the expectation from equation~\ref{ratio_op}. It may indicate that Hua and Yang's model works better at intermediate $p_T$ where quark coalescence dominates the particle production.

One interesting feature from the comparison of AMPT model calculation with data is that the turning  down point of $\Omega$/$\phi$ ratio seems to move to lower $p_T$ from $\sqrt{s_{NN}}=$ 200 GeV to 11.5 GeV. The $\Omega$ and $\phi$ only contain strange quark without any light quarks. Thus they are not affected by the light shower partons at high \pt, and its production will be dominated by the thermal strange quark recombination. In the AMPT model with the string melting version, the coalescence model is involved where the hadrons combine from the deconfined quarks after sufficient parton interaction. The assumption from Hwa and Yang~\cite{ratio} is consistent with the coalescence model in the AMPT model. While for the default AMPT model, the quark combined with the parent string is absent of sufficient interaction from partonic stage, thus the turning over point of the ratio being smaller.

    \begin{figure}[htbp]
      \includegraphics[width=0.52\textwidth]{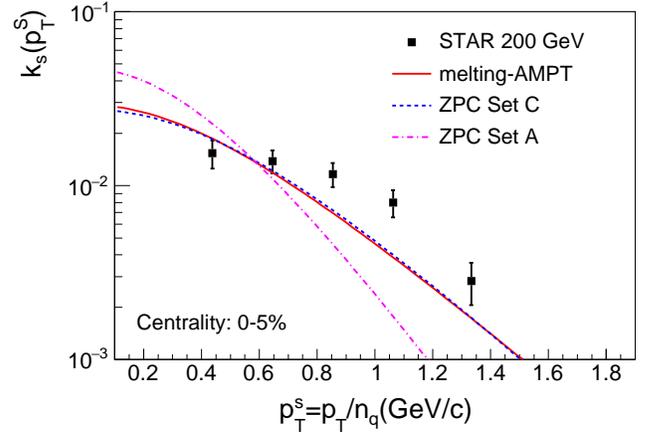}\\
      \caption{The NCQ-scaled $N(\Omega ^++\Omega^-)/[2N(\phi)]$ ratio ($k_{s}(p_{T}^{s})$)  at midrapidity ($|y|<0.5$) as a function of \pt~per constituent quark for central Au + Au collisions at $\sqrt{s_{NN}} = 200$ GeV. Solid line represents the AMPT model result. Dash lines are the strange quark \pt~distribution in AMPT from the ZPC. Experimental data~\cite{OmegaphiSTAR} is also plotted for comparison.}
      \label{zpc}
    \end{figure}

In order to investigate strangeness dynamics in more details, we compare the strange quark transverse momentum distribution after partonic interaction ceases in the AMPT model with the ratio of $\frac{N(\Omega^{-}+\Omega^{+})|_{p_{T}^{\Omega} = 3p_{T}^{s}}}{2N(\phi)|p_{T}^{\phi} = 2p_{T}^{s}}$ from final state hadrons. In the framework of quark coalescence/recombination, the $\Omega$ baryons are formed from coalescence of three strange quarks of approximately equal momentum and the $\phi$ meson is from a strange quark and an anti-strange quark. Under this logic, the production of $\Omega$ baryons is proportional to the local strange quark density, $f_{s}^{3}(p_{T}^{s})$, and the yields of $\phi$ meson is proportional to $f_{s}(p_{T}^{s})f_{\bar{s}}(p_{T}^{\bar{s}})$, where $f_{s}(f_{\bar{s}})$ is the strange (anti-strange) quark \pt~distribution at hadronization. If we take approximation of $f_{s}$ equal to $f_{\bar{s}}$, then the ratio of $\frac{N(\Omega^{-}+\Omega^{+})|_{p_{T}^{\Omega} = 3p_{T}^{s}}}{2N(\phi)|p_{T}^{\phi} = 2p_{T}^{s}}$ is proportional to $f_{s}(p_{T}^{s})$. Figure~\ref{zpc} shows the NCQ-scaled $\frac{N(\Omega^{-}+\Omega^{+})|_{p_{T}^{\Omega} = 3p_{T}^{s}}}{2N(\phi)|p_{T}^{\phi} = 2p_{T}^{s}}$ ratios and the production of strange quarks at hadronization (with scaled by an appropriate factor) as a function of $p_{T}^{s} = p_{T}/n_{q}$ at mid-rapidity ($|y|<0.5$) from central Au + Au collision at $\sqrt{s_{NN}} = 200$ GeV. The NCQ-scaled ratio from the AMPT model describes the experimental data reasonably well. The production of strange quark at hadronization after scaled by an appropriate factor is consistent with the NCQ-scaled ratio, which is a further confirmation of the coalescence assumption. This consistence also reflects the small influence of the production of $\Omega$ and $\phi$ during final state interaction of hadrons by comparing the distribution from melting-AMPT model with ZPC model of Set C. In this way, the strange quark \pt~distribution at hadronization from the parameter set C gives an appropriate description on the experimental data in contrast to the parameter set A. Therefore the parameter set C of AMPT model provides a good description of the strange quark production in Au + Au collisions at $\sqrt{s_{NN}} = 200$ GeV.

\subsection{Nuclear modification factor}

\begin{figure}[htbp]
\includegraphics[width=0.52\textwidth]{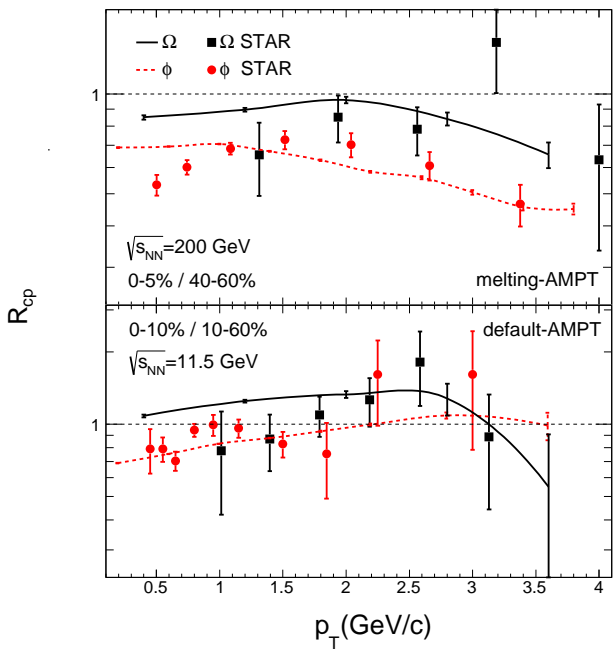}\\
\caption{The nuclear modification factor ($R_{CP}$) of $\Omega$ and $\phi$ as a function of \pt~in Au+Au collisions. (a) the AMPT model calculation with the string melting version at $\sqrt{s_{NN}} =  200$ GeV (Parameter Set C) and (b) the default AMPT model calculation at $\sqrt{s_{NN}} = 11.5$ GeV (Parameter Set A).  The experimental data~\cite{OmegaphiSTAR,star_phi} from the STAR Collaboration are plotted for comparison. }
\label{RCP}
\end{figure}

The number of binary nucleon-nucleon collision ($N_{bin}$) scaled centrality ratio of \pt~spectrum is a measure of the particle production's dependence on the collisions system's size and density \cite{STAR_Rcp}:
    \begin{equation}
        R_{CP}=\frac{[dN/dp_{T}/N_{bin}]^{Central}} {[dN/dp_{T}/N_{bin}]^{Peripheral}},
        \label{eq10}
    \end{equation}
 where $R_{CP}=1$ if particle production is equivalent to a superposition of independent nucleon-nucleon collisions, and other values illustrate for medium effect. Figure~\ref{RCP} presents the $R_{CP}$ result in the AMPT model. The collision centrality interval in $\sqrt{s_{NN}}$ = 200 GeV and 11.5 GeV is chosen differently to match the experimental data at each energy~\cite{OmegaphiSTAR,star_phi}. At $\sqrt{s_{NN}}$ = 200 GeV, the $R_{CP}$ values of both $\phi$ and $\Omega$ are all smaller than 1 and suppressed at high~\pt. At $\sqrt{s_{NN}}$ = 11.5 GeV, the $R_{CP}$ of $\phi$ meson is close to 1 and  the $R_{CP}(p_T)$ of $\Omega$ is similar to the $\phi$'s though with large statistical uncertainty. The large centrality interval of peripheral collisions (10-60\%) at $\sqrt{s_{NN}}$ = 11.5 GeV is mainly due to the limited statistics and small cross section for $\Omega$ baryon. In comparison with the data, the AMPT model calculation at $\sqrt{s_{NN}}$ = 200 GeV gives a right trend of suppression at higher \pt~even though the quantity is underestimated for $\phi$-mesons, which needs to be further improved to describe whole centrality dependence. For 11.5 GeV, our model calculations can essentially describe the data well.

\subsection{Elliptic flow of $\Omega$ and $\phi$}

We further study the elliptic flow of $\Omega$ and $\phi$ with our new tuned parameter. In previous work, we  learn that the AMPT model calculation in Au + Au collisions at $\sqrt{s_{NN}} = 200$ GeV with the Set A parameter describes the charged particle elliptic flow well at intermediate $p_T$ but overestimated the magnitude of 25\% in the low $p_T$ region~\cite{AMPT-phi-v2}.

Elliptic flow is one of the important probes of collective dynamics in heavy ion reactions \cite{probe}. It results from the spatial asymmetry in the transverse plane in non-central collisions, which is larger at early times. Therefore, the elliptic flow is sensitive to the properties of dense matter, such as the cross section of partons produced in collisions \cite{cs1,cs2,cs3,cs4} or its equation of state \cite{EOS1,EOS2,EOS3,EOS4}. Measurements of elliptic flow in Au + Au collisions at RHIC  give us much knowledge of the partonic interaction strength and the effective energy loss of partons \cite{flow_200_Omega,flow_200_phi,flow_115}.
    \begin{figure}[htbp]
      \includegraphics[width=0.52\textwidth]{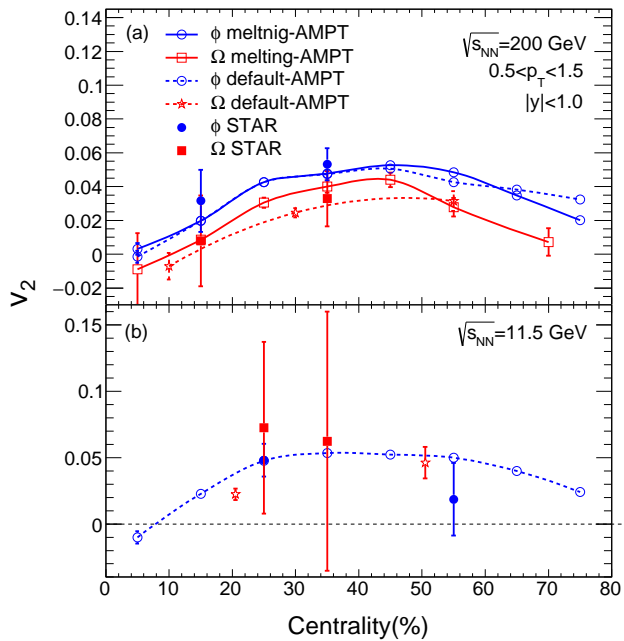}\\
      \caption{Elliptic flow of $\Omega$ and $\phi$ versus collision centrality in Au + Au collisions at RHIC. (a) the AMPT model calculation at $\sqrt{s_{NN}} =  200$ GeV (Parameter Set C) and (b) the default AMPT model calculation at $\sqrt{s_{NN}} = 11.5$ GeV (Parameter Set A). Experimental data are also plotted for comparison~\cite{flow_200_Omega,flow_200_phi,flow_115}.}
      \label{flow}
    \end{figure}

Figure \ref{flow}(a) shows the elliptic flow of $\Omega$ and $\phi$ in Au + Au collisions at $\sqrt{s_{NN}} =$ 200 GeV in broad centrality intervals. Elliptic flow of $\phi$ increases from central collision to peripheral collision up to $40\%-50\%$ and then decreases to $50\%-80\%$. The centrality dependence of the $\Omega$ elliptic flow from the AMPT model is similar to the $\phi$'s. In comparison with the data, the AMPT model with new tuned parameter perfectly describes the low \pt~$\Omega$ and $\phi$ $v_2$ both magnitude and centrality dependence. The difference between default and melting AMPT is small and future accurate data will help to distinguish the calculation. Figure \ref{flow}(b) presents the elliptic flow from the default AMPT model at $\sqrt{s_{NN}} = 11.5$ GeV. Because of the lower rate of strangeness production at $\sqrt{s_{NN}} = 11.5$ GeV, the statistical error is large both from simulation and experimental data.

\section{Summary}

Multi-strange particles, namely $\phi$ and $\Omega$, are studied in Au + Au collisions at RHIC in a multiphase transport model. Firstly, a new set of parameters (Set C) is adopted in the AMPT model which provides us a relative accurate description to the production of  $\pi$ and $\bar{p}$ invariant yields  at $\sqrt{s_{NN}} = 200$ GeV.  Secondly, the transverse momentum spectra of $\Omega$ and $\phi$ are presented. The AMPT model with the string melting scenario describes the experimental data well at $\sqrt{s_{NN}} = 200$ GeV, while the default AMPT model provides a good description at $\sqrt{s_{NN}} = 11.5$ GeV. However, we need to keep caution that the $\Omega$ yield is still underestimated by a factor 5 even though the  new set of parameters is  used, which remain space to future improvement of the model or coalescence mechanism. The parton interaction is dominant in Au + Au collision at $\sqrt{s_{NN}}$ = 200 GeV and the interaction strength becomes weak in collisions at $\sqrt{s_{NN}}$ = 11.5 GeV. The argument is further supported by the study on the ratio of $N(\Omega ^++\Omega^-)/[2N(\phi)]$ versus transverse momentum. The deconfined quarks are involved at $\sqrt{s_{NN}} = 200$ GeV in the melting AMPT model, which leads to a linear increase of the ratio, and is in a good agreement with other recombination model calculation~\cite{ratio}. The $N(\Omega ^++\Omega^-)/[2N(\phi)]$ ratio is proportional to the strange quark density which provides another support of the assumption of coalescence model. The value of the $N(\Omega ^++\Omega^-)/[2N(\phi)]$ ratio turns down at lower $p_T$ from $\sqrt{s_{NN}}$ =  200 GeV to 11.5 GeV, which indicates the transition from partonic dominated to hadronic dominated dynamics as the data displays \cite{OmegaphiSTAR}.
 Our calculations also give reasonable description to the nuclear modification factor data for Au + Au collisions at $\sqrt{s_{NN}}$ = 200 GeV and 11.5 GeV using the melting AMPT and default AMPT, respectively.
Elliptic flows of $\Omega$ and $\phi$ can describe the experimental data well at low \pt~with the new parameter set. In the future, the STAR experiment will accumulate large datasets for Au + Au collisions over a range of beam energies during 2019 $\&$ 2020 Beam Energy Scan Phase-II runs, which will further improve the precision on the $\Omega$ and $\phi$ measurements and will likely provide new insight into the effective degree of freedom of the system created at RHIC.

\section*{Acknowledgments}
This work was supported in part by the National Natural Science Foundation of China under contract Nos. 11421505, 11520101004, 11220101005, 11275250 and 11322547,  the Major State Basic Research Development Program in China under Contract No. 2014CB845400 and No. 2015CB856904, and the CAS Project Grant No. QYZDJSSW-SLH002.

\end{CJK*}

\end{document}